\documentclass[prl,twocolumn,aps,showpacs]{revtex4-1}

\usepackage{amsmath}
\usepackage{amssymb}
\usepackage{graphicx}
\usepackage{epstopdf}
\usepackage[colorlinks=true]{hyperref}

\begin{document}
\title{Spin Analogues of Superconductivity and the Integer Quantum Hall Effect\\ in an Array of Spin Chains}

\author{Daniel Hill}
\affiliation{Department of Physics and Astronomy, University of California, Los Angeles, California 90095, USA}
\author{Se Kwon Kim}
\affiliation{Department of Physics and Astronomy, University of California, Los Angeles, California 90095, USA}
\author{Yaroslav Tserkovnyak}
\affiliation{Department of Physics and Astronomy, University of California, Los Angeles, California 90095, USA}

\begin{abstract}
Motivated by the successful idea of using weakly-coupled quantum electronic wires to realize the quantum Hall effects and the quantum spin Hall effects, we theoretically construct two systems composed of weakly-coupled quantum spin chains, which can exhibit spin analogues of superconductivity and the integer quantum Hall effect. Specifically, a certain bilayer of two arrays of interacting spin chains is mapped, via the Jordan-Wigner transformation, to a negative-$U$ Hubbard model that exhibits superconductivity. In addition, an array of spin-orbit-coupled spin chains in the presence of an suitable external magnetic field is transformed to an array of quantum wires that exhibits the integer quantum Hall effect. The resultant spin superconductivity and spin integer quantum Hall effect can be characterized by their ability to transport spin without any resistance.
\end{abstract}

\date{\today}


\maketitle

\emph{Introduction.}|In a metal under normal conditions, an electric current flows in the presence of a finite resistance engendered by, e.g., scattering with impurities. The lost electrical energy due to the resistance is dissipated into heat, which is referred to as Joule heating that opposes the efficient use of the energy. There are, however, two physical phenomena under special conditions that allow an electric current to flow without any resistance. The one is superconductivity occurring at low temperatures~\cite{*[][{, and references therein.}] Tinkham2004}. Its first microscopic theory was given in 1957 by Bardeen, Cooper, and Schrieffer~\cite{BardeenPR1957}, who showed that superconductivity can be understood as a property of macroscopic quantum wavefunction of condensed pairs of electrons subsequently termed Cooper pairs. The other is the set of quantum Hall effects exhibited in two-dimensional systems at low temperatures and strong magnetic fields~\cite{*[][{, and references therein.}] Prange1990}. The integer quantum Hall effect is the first of such that was discovered in 1980 by \textcite{KlitzingPRL1980}. It occurs when the number of electrons per unit magnetic flux takes an integer value $\nu$, leading to the situation in which the bulk is gapped, but the edge supports $\nu$ gapless modes with no resistance.

Spintronics aims at harnessing the spin degrees of freedom to advance from conventional charge-based electronics~\cite{WolfScience2001, *ZuticRMP2004}. In particular, magnetic insulators that are free from Joule heating have been gaining attention in the field owing to their potential advantage of low-energy consumption. An efficient spin transport in such magnetic insulators is one of the important topics in spintronics, and researchers have been investigating possible ways to achieve it by borrowing some ideas from the aforementioned phenomena of dissipationless charge transport. For example, a spin-analogue of an electric supercurrent supported in easy-plane magnets has been theoretically investigated~\cite{HalperinPR1969, *SoninJETP1978, *SoninAP2010, *ChenPRB2014-2, *ChenPRB2014, *TakeiPRL2014, *TakeiPRB2014, *ChenPRL2015}, which is shown to decay algebraically as a function of the distance from the spin-injection point contrary to an exponential decay of a diffusive spin current. Spin analogues of the integer and fractional quantum Hall phases have also been put forward in the studies of spin liquids~\cite{KalmeyerPRL1987, *KalmeyerPRB1989, *WenPRB1989, *MengPRB2015, HaldanePRB1995} and topological magnon insulators~\cite{MatsumotoPRL2011, *MatsumotoPRB2012, *ShindouPRB2013, *ShindouPRB2013-2, *ZhangPRB2013, *MookPRB2014, *MookPRB2014-2}.

\begin{figure}
\includegraphics[width=\columnwidth]{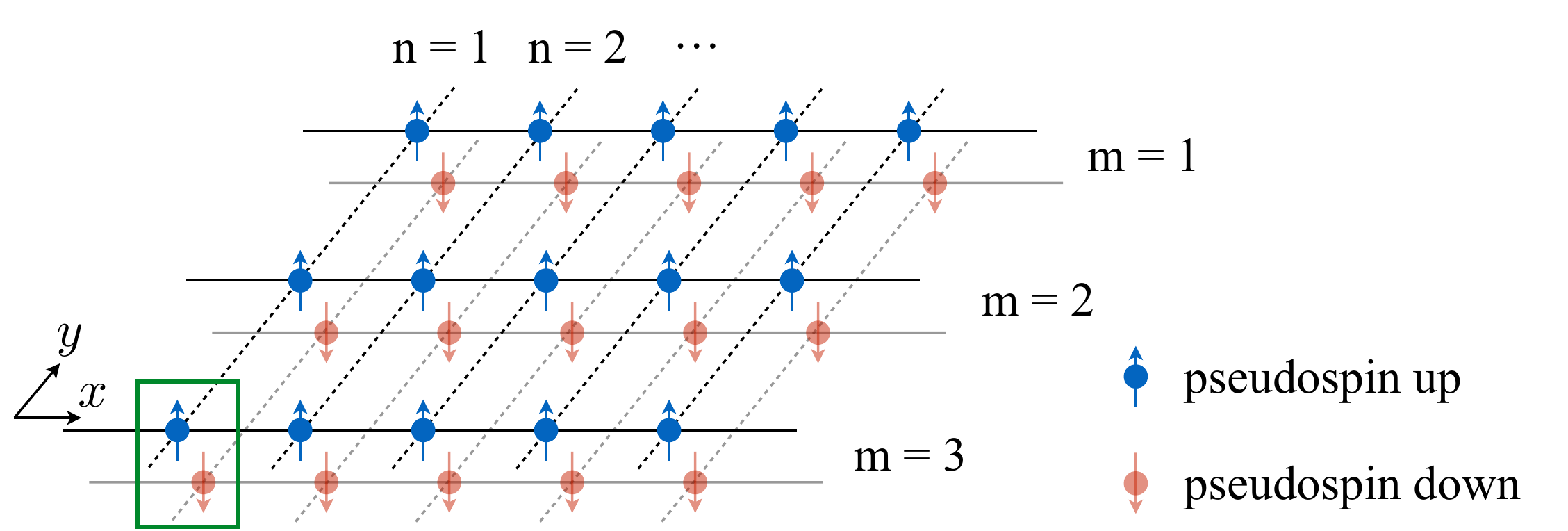}
\caption{Schematic of a bilayer of two arrays of weakly-coupled spin chains (shown as the solid lines indexed by $m$), each of which can be represented by a one-dimensional system of (spinless) Jordan-Wigner fermions. The top and bottom layer indices serve as the pseudospin up and down for the fermions, respectively. The green box represents a pseudospin-singlet Cooper pair of two fermions established by an Ising interlayer interaction.}
\label{fig:fig1}
\end{figure}

\begin{figure}
\includegraphics[width=\columnwidth]{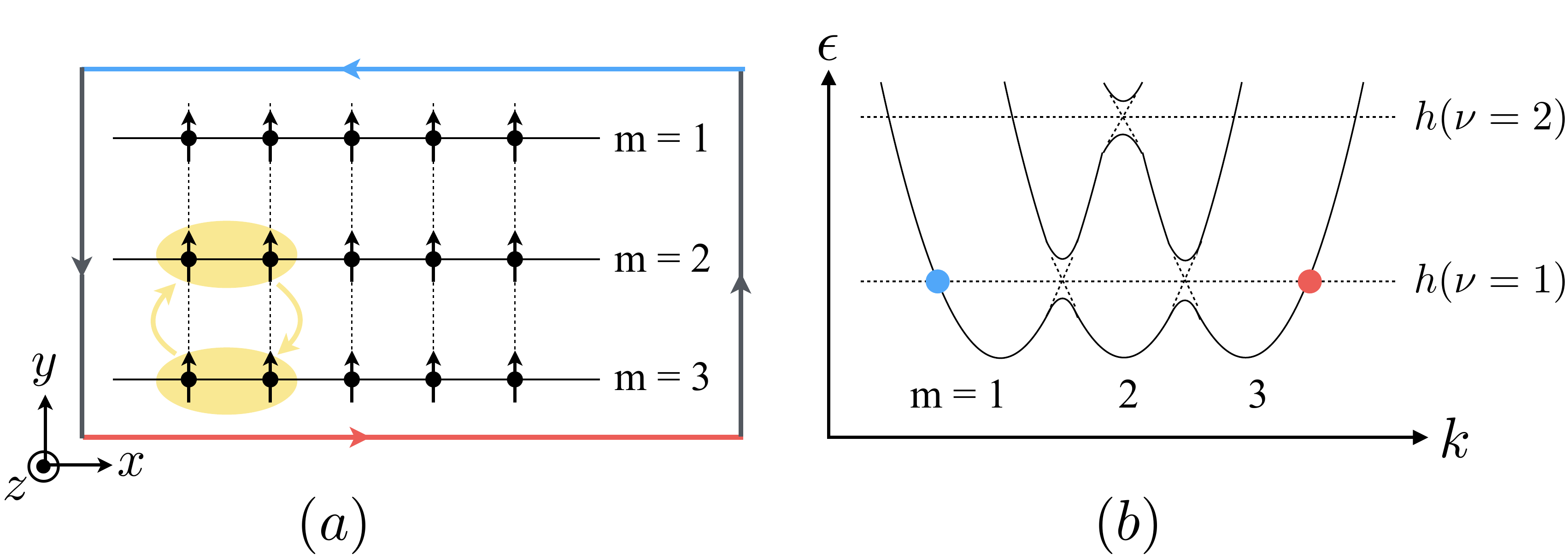}
\caption{(a) Schematic of an array of spin-orbit-coupled spin-1/2 spin chains, which can support chiral edge modes of the Jordan-Wigner fermions. The coupling of four spins (colored by yellow) illustrates the interchain interaction $\mathcal{O}$ [Eq.~(\ref{eq:O})]. (b) A schematic plot showing how the interchain interaction gives rise to chiral edge modes with the gapped bulk. At the external magnetic field $h$ corresponding to the filling factor $\nu = 1$, the JW fermion can flow in the left direction on the top chain (colored by blue) and in the right direction on the bottom chain (colored by red) in (a), which are represented by the left blue and right red dots in (b), respectively. The particle current in the JW representation corresponds to the spin current polarized along the $z$ axis.}
\label{fig:fig2}
\end{figure}

In this Letter, we theoretically construct two spin systems, which can exhibit spin analogues of superconductivity and the integer quantum Hall effect, by using weakly-coupled quantum spin chains. Our work is motivated by the successful theoretical realizations of the quantum Hall phases and the quantum spin Hall phases in an array of quantum electronic wires~\cite{YakovenkoPRB1991, *KlinovajaPRL2013, KanePRL2002, *KlinovajaPRB2014}. Specifically, first, we show that an Ising-coupled bilayer of two arrays of weakly-coupled quantum XX spin chains can be mapped to a negative-$U$ Hubbard model for electrons by the Jordan-Wigner (JW) transformation~\cite{JordanZP1928, GalitskiPRB2010} within a mean-field treatment of the interchain coupling. Since the particle current in the JW representation corresponds to the spin current polarized along the z axis, the established charge superconductivity of the negative-$U$ Hubbard model~\cite{*[][{, and references therein.}] MicnasRMP1990} naturally translates into spin superconductivity of our original spin system. See Fig.~\ref{fig:fig1} for an illustration of the system. Secondly, we show that an array of weakly-coupled quantum XX spin chains with Dzyaloshinskii-Moriya (DM) intrachain interaction can be transformed to an array of quantum electronic wires subjected to an external magnetic field by the same approach taken for spin superconductivity. The integer quantum Hall effect of the latter electronic system~\cite{KanePRL2002} then translates into its spin analogue of the former spin system. See Fig.~\ref{fig:fig2}(a) for an illustration of the system. The only nontrivial analytical effort in this work, besides performing the JW transformation, is the mean-field calculation of the interchain hopping terms.

\emph{Main results.}|Our main results can be summarized as follows. First, for spin superconductivity, we consider the following spin Hamiltonian for two layers of weakly-coupled $M$ spin-1/2 chains of length $N$:
\begin{equation}
\label{eq:H-sc}
\begin{split}
H_\text{sc} = &  J \sum_{n, m, \alpha} \boldsymbol{\sigma}^{\parallel}_{n, m, \alpha} \cdot \boldsymbol{\sigma}^{\parallel}_{n+1, m, \alpha} \\ 
& - H \sum_{n, m, \alpha} \sigma_{n, m, \alpha}^z - U \sum_{n, m} \sigma^z_{n, m, \uparrow} \sigma^z_{n, m, \downarrow} \\
& - K \sum_{n, m, \alpha} \left[ \mathcal{O}_{n, m, \alpha} + \text{H.c.} \right] \, , \\
\end{split}
\end{equation}
with
\begin{equation}
\label{eq:O}
\mathcal{O}_{n, m, \alpha} = \sigma^+_{n, m, \alpha} \sigma^+_{n+1, m, \alpha} \sigma^-_{n, m+1, \alpha} \sigma^-_{n+1, m+1, \alpha} \, ,
\end{equation}
where the integers $m$ and $n$ are the indices for a spin chain within a layer and a spin within a chain, respectively, and $\alpha = \uparrow, \downarrow$ indexes the layer which will serve as the pseudospin of the JW fermions. A spin is represented by the three-dimensional Pauli matrices $\boldsymbol{\sigma}$; the symbol $\parallel$ denotes the projection of the vector onto the $xy$ plane; $\sigma^\pm \equiv (\sigma^x \pm i \sigma^y) / 2$. Here, the first term describes the quantum antiferromagnetic XX spin-1/2 chains with $J > 0$~\footnote{We do not lose any generality by choosing the sign of the exchange interaction, $J > 0$ here, for the sign can be flipped by rotating every other spin by $\pi$ around the $z$ axis.}; the second term is the Zeeman energy; the third term is the ferromagnetic Ising interaction between the two layers; the last term represents a weak four-spin interaction with $0 < K \ll J$, which, in the JW representation, can engender the interchain tunneling and thereby make each layer an effective two-dimensional fermionic gas. Interchain interactions involving only two spins such as the Heisenberg XX exchange $\propto \boldsymbol{\sigma}_{n, m}^\parallel \cdot \boldsymbol{\sigma}_{n, m+1}^\parallel$ would also appear as tunneling between two chains. They, however, introduce nonlocal terms after the JW transformation, making it difficult to treat the interchain interaction~\footnote{The long-range interchain interactions may be treated within the framework of the coupled Luttinger liquids~\cite{*[][{, and references therein.}] GiamarchiCR2004}, but it is beyond the scope of our work.}. Our goal, instead, is to construct simple spin systems that can be viewed as weakly-interacting simple fermionic wires. Therefore, by coupling neighboring spin chains by the four-spin interaction, we retain its locality after the JW transformation. The spin Hamiltonian $H_\text{sc}$ respects the spin-rotational symmetry about the $z$ axis, and thus the total spin projected onto the same axis is conserved. 

The spin Hamiltonian $H_\text{sc}$ can be transformed into the Hamiltonian for the spinless fermions by the multi-dimensional JW transformation~\cite{GalitskiPRB2010}:
$
f_{n, m, \alpha} = \sigma^-_{n, m, \alpha} \left(\prod_{l < n} \sigma^z_{l, m, \alpha} \right) \left( \prod_{(k, \beta) < (m, \alpha)} \tau^y_{k, \beta} \right) \tau^x_{m, \alpha} \, ,
$
and the analogous expression for $f^\dagger_{n, m, \alpha}$ with $\sigma^-$ substituted by $\sigma^+$, where the auxiliary Pauli-matrix vector, $\boldsymbol{\tau}_{m, \alpha}$, is introduced for each spin chain to make the fermion operators on different chains anticommute~\footnote{The comparison of two tuples, $(k, \beta)$ and $(m, \alpha)$, representing two spin chains is determined by their orders in a serialization of spin chains. One can take an alternative JW transformation by restricting $\beta$ to $\alpha$ in the product and introducing a new additional Pauli-matrix vector for maintaining anticommutation between the two layers.}. The interchain interaction yields a quartic term in the fermion operators, and thus we take the mean-field approach to study its effect. We will show that the resultant mean-field Hamiltonian is given by the attractive Hubbard model~\cite{MicnasRMP1990}:
\begin{equation}
\begin{split}
\bar{H}_\text{sc} = & - t_x \sum_{n, m, \alpha} \left[ f^\dagger_{n, m, \alpha} f_{n+1, m, \alpha} + \text{H.c.} \right] \\
& - t_y \sum_{n, m, \alpha} \left[ f^\dagger_{n, m, \alpha} f_{n, m+1, \alpha} + \text{H.c.} \right] \\
& - \mu \sum_{n, m, \alpha} n_{n, m, \alpha} - u \sum_{n, m} n_{n, m, \uparrow} n_{n, m, \downarrow} \, ,
\end{split}
\end{equation}
where $n_{n, m, \alpha} = f_{n, m, \alpha}^\dagger f_{n, m, \alpha}$ is the fermion-number operator, $t_x = 2 J, t_y = 2 K \chi, \mu = 2 H - 2 U$, and $u = 4 U$. Here, $\chi \equiv \langle \sum_{n, m, \alpha} f_{n, m, \alpha}^\dagger f_{n, m+1, \alpha} \rangle / 2NM$ is the mean field for the interchain tunneling~\footnote{The Hartree-Fock decoupling is taken only for the interchain interaction, not for the Ising interlayer interaction.}. By $\bar{H}$, we will denote the Hamiltonians in the JW representation throughout. When the Fermi energy lies close to the bottom of the band for a single chain, $\mu = - 2 t_x + \delta \mu$ with $|\delta \mu| \ll K$, the interchain tunneling amplitude is given by $\chi \approx K / 25 t_x$. From the results known for the attractive Hubbard model~\cite{MicnasRMP1990}, we can conclude that the ground state of the Hamiltonian $\bar{H}_\text{sc}$ away from the half-filling is in the superconducting phase composed of pseudospin-singlet Cooper pairs of the JW fermions, which should exhibit a spin-analogue of charge superconductivity~\footnote{In the weak-attraction limit, $u \ll t_x \, , t_y$, the superconducting gap is given by $\Delta \sim t \sqrt{\rho (2 - \rho)} / \sinh(t / u)$, where $t \equiv t_x + t_y$ is the bandwidth and $\rho$ is the number of electrons per site~\cite{MicnasRMP1990}.}. 

Second, for a spin analogue of the integer quantum Hall effect, we take the following spin Hamiltonian:
\begin{equation}
\label{eq:H-qh}
\begin{split}
H_\text{qh} = &  J \sum_{n, m} \cos(m \phi) \, \boldsymbol{\sigma}_{n, m}^\parallel \cdot \boldsymbol{\sigma}_{n+1, m}^\parallel  \\
& + J \sum_{n, m} \sin(m \phi) \,  \hat{\mathbf{z}} \cdot \boldsymbol{\sigma}_{n, m} \times \boldsymbol{\sigma}_{n+1, m} \\
& - H \sum_{n, m} \sigma_{n, m}^z - K \sum_{n, m} \left[ \mathcal{O}_{n, m} + \text{H.c.} \right] \, ,
\end{split}
\end{equation}
where the four-spin interaction $\mathcal{O}_{n, m}$ is given by Eq.~(\ref{eq:O}) with $\alpha$ removed. Here, the first two terms describe the antiferromagnetic Heisenberg XY spin chains with the DM interaction; the third term is the Zeeman coupling; the last term is the weak interchain interaction, $0 < K \ll J$. See Fig.~\ref{fig:fig2}(a) for an illustration of the system. The DM interaction can exist if the reflection symmetry with respect to the $xz$ plane is broken; the Hamiltonian respects the reflection symmetries through the $xy$ and $yz$ planes. The chain-dependent exchange coefficients can be realized by controlling the extent of the reflection-symmetry breaking associated with the DM interaction. We focus on the weak DM interactions, $0 < \phi J \ll K$, comparing to the interchain coupling~\footnote{For the strong DM interactions comparing to the interchain coupling, $0 < K \ll \phi J$, the mean field for the interchain tunneling is exponentially small, $\chi \sim \exp(- 4 \pi \phi J / K)$.}.

By employing the JW transformation~\cite{GalitskiPRB2010} and taking the mean-field approach for the interchain interaction, as shown below, we obtain the following tight-binding Hamiltonian:
\begin{equation}
\begin{split}
\bar{H}_{\text{qh}} = & - t_x \sum_{n, m} \left[ e^{i m \phi} f^\dagger_{n, m} f_{n+1, m} + \text{H.c.} \right] \\
& - t_y \sum_{n, m} \left[ f^\dagger_{n, m} f_{n, m+1} + \text{H.c.} \right] \\
& - \mu \sum_{n, m} n_{n, m} \, ,
\end{split}
\end{equation}
which describes an array of quantum electronic wires in the presence of an external magnetic field $\propto \phi$. The parameters are given by $t_x = 2 J, t_y = 2 K \chi$, and $\mu = 2 H$ with the mean-field interchain tunneling $\chi$. The integer quantum Hall effect at the filling factor $\nu = 1$ arises when the Fermi energy is close to the crossing point of the two bands of adjacent chains, $\mu = - t_x (2 + \phi^2) + \delta \mu$ with $|\delta \mu| \ll t_x$. The self-consistent solution is then given by $\chi \approx K / 25 t_x$. The integer quantum Hall effects at higher filling factors $\nu$ can be analogously obtained in the $\nu$th order of the perturbative treatment of the interchain interaction~\cite{KanePRL2002, KlinovajaPRB2014}.

The Hamiltonian $\bar{H}_\text{qh}$ has been shown to exhibit the integer quantum Hall effect~\cite{YakovenkoPRB1991, KanePRL2002}. Let us briefly explain how the integer quantum Hall effects arise in the model for an example of filling factor $\nu = 1$. See Fig.~\ref{fig:fig2}(b) for the JW fermion bands of spin chains and the gap openings by the interchain tunneling. When the Fermi energy $\mu$ lies in the bulk gap, there are one gapless mode in the top chain ($m = 1$ in the figure) and the other in the bottom chain ($m = 3$ in the figure). The two modes propagate in the opposite directions, and thus engender one chiral edge mode together. The integer quantum Hall effect at higher filling factors $\nu$ supports $\nu$ chiral edge modes by an analogous mechanism~\cite{KanePRL2002, KlinovajaPRB2014}. The state we obtained is different from the conventional quantum Hall phase in that the transported quantity is spin, not charge; it is also distinct from the traditional quantum spin Hall phase~\cite{QiRMP2011} in that the resultant spin transport does not accompany any charge transport.

Although, to the best of our knowledge, there is no physical system that can realize our proposal, let us make some comments about experimental realizations. First, spin-1/2 chain systems Cs$_2$CoCl$_4$~\cite{*[][{, and references therein.}] BreunigPRL2013} and PrCl$_3$~\cite{*[][{, and references therein.}] GoovaertsPRL1984} are known to be well described by the isotropic Heisenberg XX model. Secondly, the DM interaction in a single chain can be induced by breaking the reflection symmetry through the $xz$ plane, which can be, in principle, realized by attaching a one-dimensional nonmagnetic material next to the spin chain. The gradient in the DM interaction can be engendered by modulating the distance between the nonmagnetic material and the spin chain. Lastly, the four-spin exchange interaction can arise as the fourth-order term in the strong-coupling expansion of the half-filled Hubbard model or due to the spin-lattice coupling, and its magnitude can be comparable to two-spin Heisenberg exchange in certain materials~\cite{Mikeska2004, GritsevPRB2004, *CapponiPRB2013}.

\emph{Spin superconductivity.}|We explain how we obtained the mean-field values for the interchain tunneling amplitude for the case of spin superconductivity. To analyze the effects of the quartic fermion operator $f^\dagger_{n, m, \alpha} f^\dagger_{n+1, m, \alpha} f_{n, m+1, \alpha} f_{n+1, m+1, \alpha}$ mapped from $\mathcal{O}_{n, m, \alpha}$ [Eq.~(\ref{eq:O})], we employ the Hartree-Fock decoupling~\cite{Bruus}. There are two potentially relevant mean-field order parameters~\footnote{There is the third mean field, $\sum_{n, m, \alpha} \protect\langle f^\dagger_{n, m, \alpha} f_{n+1, m+1, \alpha} \protect\rangle$, but it is omitted from the analysis because its effects are similar to those of $\chi$ and thus do not change the qualitative property of the system.}. One is an interchain-tunneling amplitude,
\begin{equation}
\chi = \frac{1}{2NM} \sum_{n, m, \alpha} \langle f^\dagger_{n, m, \alpha} f_{n, m+1, \alpha} \rangle \, .
\end{equation}
The other is an intrachain Cooper-pairing amplitude, 
\begin{equation}
\Delta = \frac{1}{2NM} \sum_{n, m, \alpha} \langle f_{n, m, \alpha} f_{n+1, m, \alpha} \rangle \, .
\end{equation}
The mean-field Hamiltonian for a single layer is given by
\begin{equation}
\label{eq:H-alpha}
\begin{split}
\bar{H}_{\alpha} = & - t_x \sum_{n, m} \left[ f^\dagger_{n, m, \alpha} f_{n + 1, m, \alpha} + \text{H.c.} \right] \\
& - 2K \sum_{n, m} \left[ \chi f^\dagger_{n, m, \alpha} f_{n, m + 1, \alpha} + \text{H.c.} \right] \\
& - 2K \sum_{n, m} \left[ \Delta f^\dagger_{n, m, \alpha} f^\dagger_{n+1, m, \alpha} + \text{H.c.} \right] \\
& - \mu \sum_{n, m} n_{n, m, \alpha} \, ,
\end{split}
\end{equation}
up to an additive constant, in the limit of zero interlayer coupling $U \rightarrow 0$. Assuming the periodic boundary conditions, the self-consistency equations for the two mean-field order parameters $\chi$ and $\Delta$ in the momentum space are given by
\begin{eqnarray}
\Delta &=& \frac{1}{NM} \sum_{\mathbf{k}} \frac{2K \Delta \sin^2 k_x}{\sqrt{\epsilon(\mathbf{k})^2 + |\Delta(\mathbf{k})|^2}} \, , \\
\chi &=& \frac{1}{NM} \sum_\mathbf{k} \frac{\cos k_y}{2} \left( 1 - \frac{\epsilon(\mathbf{k})}{\sqrt{\epsilon(\mathbf{k})^2 + |\Delta(\mathbf{k})|^2}} \right) \, ,
\end{eqnarray}
where $\epsilon(\mathbf{k}) = - 2 t_x \cos k_x - 4 K \chi \cos k_y - \mu$ and $\Delta(\mathbf{k}) = 4 i K \Delta \sin k_x$. Here, the spatial coordinates $x$ and $y$ are the continuum analogues of $n$ and $m$. Since the coefficient for the interchain interaction is assumed to be positive, $K > 0$, the Cooper-pairing amplitude vanishes, $\Delta = 0$. To compute the self-consistent solution for $\chi$ analytically, we assume that the effective chemical potential is closed to the bottom of the band for a single chain, $\mu = - 2 t_x + \delta \mu$ with $|\delta \mu| \ll K$, and use a parabolic band approximation for the dispersion $\epsilon(\mathbf{k})$ around the origin. We then obtain
\begin{equation} 
\label{eq:chi}
\chi = \left( \int_0^{\pi/2} dk_y \, \cos^{3/2} k_y \right)^2 \frac{4 K}{\pi^4 t_x} + \text{O} \left( \frac{\delta \mu}{K} \right) \, ,
\end{equation}
which can be approximated to $\chi \approx K / 25 t_x$. With the finite $\chi_\alpha = \chi$ and vanishing $\Delta_\alpha = 0$, the mean-field Hamiltonian $\bar{H}_\alpha$~(\ref{eq:H-alpha}) for a single layer describes a two-dimensional spinless fermion gas.

\emph{Spin integer quantum Hall effect.}|Next, we explain the derivation of the mean-field results for the case of spin integer quantum Hall effect. Since the gradient of spin-orbit coupling breaks the translational symmetry of the system along the $y$ axis, it is difficult to obtain an analytical mean-field solution $\chi$ for arbitrary $M$. Instead, let us consider a special case of two weakly-coupled spin chains, which is described by $H_\text{qh}$ [Eq.~(\ref{eq:H-qh})] with $m = \pm 1$. Two possible order parameters pertain to the interchain tunneling, $\chi = \sum_{n} \langle f^\dagger_{n, 1} f_{n, -1} \rangle / N$, and the Cooper pairing, $\Delta = \sum_{n, m = \pm 1} \langle f_{n, m} f_{n +1, m} \rangle / 2 N$. The mean-field Hamiltonian for the JW fermions in the momentum space is given by
\begin{equation}
\begin{split}
\bar{H} = & \sum_{k, m = \pm 1} \left[ ( - 2 t_x \cos(k + m \phi) - \mu ) f^\dagger_{k, m} f_{k, m} \right] \\
& - 2 K \sum_{k} \left[ \chi f^\dagger_{k, 1} f_{k, -1} + \text{H.c.} \right] \\
& - 2 K \sum_{k, m = \pm 1} \left[ \Delta e^{i k } f^\dagger_{k, m} f^\dagger_{-k, m} + \text{H.c.} \right] \, .
\end{split}
\end{equation}

We will assume that two phases with finite $\chi$ and $\Delta$ are mutually exclusive, and will treat them separately. For $K > 0$, which is assumed throughout the Letter, the self-consistency equation yields a vanishing Cooper-pairing amplitude, $\Delta = 0$, as in the case of spin superconductivity. With $\Delta = 0$, the band structure of the Hamiltonian is $\epsilon_{\pm} (k) = t_x k^2 - \delta \mu \pm 2 \sqrt{(t_x \phi k)^2 + (K \chi)^2}$ for $|k|, |\phi| \ll 1$, where $\mu =  - t_x (2 + \phi^2) + \delta \mu$. When the effective Fermi energy is at the band-crossing point, $\delta \mu = 0$, the analytical solution to the self-consistency equation for $\chi$ is given by $\chi \simeq K / 2 \pi^2 t_x$, which agrees well with $\chi \approx K / 25 t_x$ [Eq.~(\ref{eq:chi})] obtained for the $M \rightarrow \infty$ case. The finite interchain tunneling $\chi > 0$ opens up the gap at the crossing point of the two bands of chains. See Fig.~\ref{fig:fig2}(b) for illustrations of the gap openings.

\emph{Discussion.}|We have theoretically constructed the two models of an array of weakly-coupled spin chains, which can exhibit spin analogues of charge superconductivity and the integer quantum Hall effect. To drive spin current through those systems, we can apply an external-magnetic-field gradient, which acts as an electric field on the JW fermions~\cite{HaldanePRB1995}. We can also attach the spin system to heavy metals such as platinum, which can directly inject a spin current to proximate magnets via spin Hall effects~\cite{*[][{, and references therein.}] SinovaRMP2015}. Reciprocally, a spin current out of the system can be measured via inverse spin Hall effects by putting it next to heavy metals. Spin superconductivity and spin integer quantum Hall effects can be characterized by the zero resistance in spin flow through the bulk and along the boundary, respectively, when neglecting spin dissipation due to, e.g., thermal fluctuations or spin-lattice coupling.

From the results obtained for quantum spin chains, we expect that an array of weakly-coupled classical Heisenberg spin chains (that are composed of large spins) with the DM interaction in the presence of a strong external magnetic field would support the magnonic chiral edge modes by forming a topological magnon insulator~\cite{MatsumotoPRL2011} under suitable conditions. More broadly, we envision that weakly-coupled one-dimensional magnetic materials would serve as a versatile platform to engineer various spin-related topological phases.

\begin{acknowledgments}
The authors thank Pramey Upadhyaya for insightful discussions. This work was supported by the Army Research Office under Contract No. 911NF-14-1-0016.
\end{acknowledgments}

\bibliography{/Users/evol/Dropbox/School/Research/master}

%merlin.mbs apsrev4-1.bst 2010-07-25 4.21a (PWD, AO, DPC) hacked
%Control: key (0)
%Control: author (8) initials jnrlst
%Control: editor formatted (1) identically to author
%Control: production of article title (-1) disabled
%Control: page (0) single
%Control: year (1) truncated
%Control: production of eprint (0) enabled
\begin{thebibliography}{49}%
\makeatletter
\providecommand \@ifxundefined [1]{%
 \@ifx{#1\undefined}
}%
\providecommand \@ifnum [1]{%
 \ifnum #1\expandafter \@firstoftwo
 \else \expandafter \@secondoftwo
 \fi
}%
\providecommand \@ifx [1]{%
 \ifx #1\expandafter \@firstoftwo
 \else \expandafter \@secondoftwo
 \fi
}%
\providecommand \natexlab [1]{#1}%
\providecommand \enquote  [1]{``#1''}%
\providecommand \bibnamefont  [1]{#1}%
\providecommand \bibfnamefont [1]{#1}%
\providecommand \citenamefont [1]{#1}%
\providecommand \href@noop [0]{\@secondoftwo}%
\providecommand \href [0]{\begingroup \@sanitize@url \@href}%
\providecommand \@href[1]{\@@startlink{#1}\@@href}%
\providecommand \@@href[1]{\endgroup#1\@@endlink}%
\providecommand \@sanitize@url [0]{\catcode `\\12\catcode `\$12\catcode
  `\&12\catcode `\#12\catcode `\^12\catcode `\_12\catcode `\%12\relax}%
\providecommand \@@startlink[1]{}%
\providecommand \@@endlink[0]{}%
\providecommand \url  [0]{\begingroup\@sanitize@url \@url }%
\providecommand \@url [1]{\endgroup\@href {#1}{\urlprefix }}%
\providecommand \urlprefix  [0]{URL }%
\providecommand \Eprint [0]{\href }%
\providecommand \doibase [0]{http://dx.doi.org/}%
\providecommand \selectlanguage [0]{\@gobble}%
\providecommand \bibinfo  [0]{\@secondoftwo}%
\providecommand \bibfield  [0]{\@secondoftwo}%
\providecommand \translation [1]{[#1]}%
\providecommand \BibitemOpen [0]{}%
\providecommand \bibitemStop [0]{}%
\providecommand \bibitemNoStop [0]{.\EOS\space}%
\providecommand \EOS [0]{\spacefactor3000\relax}%
\providecommand \BibitemShut  [1]{\csname bibitem#1\endcsname}%
\let\auto@bib@innerbib\@empty
%</preamble>
\bibitem [{\citenamefont {Tinkham}(2004)}]{Tinkham2004}%
  \BibitemOpen
  \bibfield  {author} {\bibinfo {author} {\bibfnamefont {M.}~\bibnamefont
  {Tinkham}},\ }\href@noop {} {\emph {\bibinfo {title} {Introduction to
  Superconductivity}}}\ (\bibinfo  {publisher} {Dover, New York},\ \bibinfo
  {year} {2004})\BibitemShut {NoStop}%
\bibitem [{\citenamefont {Bardeen}\ \emph {et~al.}(1957)\citenamefont
  {Bardeen}, \citenamefont {Cooper},\ and\ \citenamefont
  {Schrieffer}}]{BardeenPR1957}%
  \BibitemOpen
  \bibfield  {author} {\bibinfo {author} {\bibfnamefont {J.}~\bibnamefont
  {Bardeen}}, \bibinfo {author} {\bibfnamefont {L.~N.}\ \bibnamefont {Cooper}},
  \ and\ \bibinfo {author} {\bibfnamefont {J.~R.}\ \bibnamefont {Schrieffer}},\
  }\href {\doibase 10.1103/PhysRev.108.1175} {\bibfield  {journal} {\bibinfo
  {journal} {Phys. Rev.}\ }\textbf {\bibinfo {volume} {108}},\ \bibinfo {pages}
  {1175} (\bibinfo {year} {1957})}\BibitemShut {NoStop}%
\bibitem [{\citenamefont {Prange}\ and\ \citenamefont
  {Girvin}(1990)}]{Prange1990}%
  \BibitemOpen
  \bibinfo {editor} {\bibfnamefont {R.~E.}\ \bibnamefont {Prange}}\ and\
  \bibinfo {editor} {\bibfnamefont {S.~M.}\ \bibnamefont {Girvin}},\ eds.,\
  \href@noop {} {\emph {\bibinfo {title} {The Quantum Hall Effect}}}\ (\bibinfo
   {publisher} {Springer, New York},\ \bibinfo {year} {1990})\BibitemShut
  {NoStop}%
\bibitem [{\citenamefont {Klitzing}\ \emph {et~al.}(1980)\citenamefont
  {Klitzing}, \citenamefont {Dorda},\ and\ \citenamefont
  {Pepper}}]{KlitzingPRL1980}%
  \BibitemOpen
  \bibfield  {author} {\bibinfo {author} {\bibfnamefont {K.~v.}\ \bibnamefont
  {Klitzing}}, \bibinfo {author} {\bibfnamefont {G.}~\bibnamefont {Dorda}}, \
  and\ \bibinfo {author} {\bibfnamefont {M.}~\bibnamefont {Pepper}},\ }\href
  {\doibase 10.1103/PhysRevLett.45.494} {\bibfield  {journal} {\bibinfo
  {journal} {Phys. Rev. Lett.}\ }\textbf {\bibinfo {volume} {45}},\ \bibinfo
  {pages} {494} (\bibinfo {year} {1980})}\BibitemShut {NoStop}%
\bibitem [{\citenamefont {Wolf}\ \emph {et~al.}(2001)\citenamefont {Wolf},
  \citenamefont {Awschalom}, \citenamefont {Buhrman}, \citenamefont {Daughton},
  \citenamefont {von Moln{\'a}r}, \citenamefont {Roukes}, \citenamefont
  {Chtchelkanova},\ and\ \citenamefont {Treger}}]{WolfScience2001}%
  \BibitemOpen
  \bibfield  {author} {\bibinfo {author} {\bibfnamefont {S.~A.}\ \bibnamefont
  {Wolf}}, \bibinfo {author} {\bibfnamefont {D.~D.}\ \bibnamefont {Awschalom}},
  \bibinfo {author} {\bibfnamefont {R.~A.}\ \bibnamefont {Buhrman}}, \bibinfo
  {author} {\bibfnamefont {J.~M.}\ \bibnamefont {Daughton}}, \bibinfo {author}
  {\bibfnamefont {S.}~\bibnamefont {von Moln{\'a}r}}, \bibinfo {author}
  {\bibfnamefont {M.~L.}\ \bibnamefont {Roukes}}, \bibinfo {author}
  {\bibfnamefont {A.~Y.}\ \bibnamefont {Chtchelkanova}}, \ and\ \bibinfo
  {author} {\bibfnamefont {D.~M.}\ \bibnamefont {Treger}},\ }\href {\doibase
  10.1126/science.1065389} {\bibfield  {journal} {\bibinfo  {journal}
  {Science}\ }\textbf {\bibinfo {volume} {294}},\ \bibinfo {pages} {1488}
  (\bibinfo {year} {2001})}\BibitemShut {NoStop}%
\bibitem [{\citenamefont {\ifmmode \check{Z}\else
  \v{Z}\fi{}uti\ifmmode~\acute{c}\else \'{c}\fi{}}\ \emph
  {et~al.}(2004)\citenamefont {\ifmmode \check{Z}\else
  \v{Z}\fi{}uti\ifmmode~\acute{c}\else \'{c}\fi{}}, \citenamefont {Fabian},\
  and\ \citenamefont {Das~Sarma}}]{ZuticRMP2004}%
  \BibitemOpen
  \bibfield  {author} {\bibinfo {author} {\bibfnamefont {I.}~\bibnamefont
  {\ifmmode \check{Z}\else \v{Z}\fi{}uti\ifmmode~\acute{c}\else \'{c}\fi{}}},
  \bibinfo {author} {\bibfnamefont {J.}~\bibnamefont {Fabian}}, \ and\ \bibinfo
  {author} {\bibfnamefont {S.}~\bibnamefont {Das~Sarma}},\ }\href {\doibase
  10.1103/RevModPhys.76.323} {\bibfield  {journal} {\bibinfo  {journal} {Rev.
  Mod. Phys.}\ }\textbf {\bibinfo {volume} {76}},\ \bibinfo {pages} {323}
  (\bibinfo {year} {2004})}\BibitemShut {NoStop}%
\bibitem [{\citenamefont {Halperin}\ and\ \citenamefont
  {Hohenberg}(1969)}]{HalperinPR1969}%
  \BibitemOpen
  \bibfield  {author} {\bibinfo {author} {\bibfnamefont {B.~I.}\ \bibnamefont
  {Halperin}}\ and\ \bibinfo {author} {\bibfnamefont {P.~C.}\ \bibnamefont
  {Hohenberg}},\ }\href {\doibase 10.1103/PhysRev.188.898} {\bibfield
  {journal} {\bibinfo  {journal} {Phys. Rev.}\ }\textbf {\bibinfo {volume}
  {188}},\ \bibinfo {pages} {898} (\bibinfo {year} {1969})}\BibitemShut
  {NoStop}%
\bibitem [{\citenamefont {Sonin}(1978)}]{SoninJETP1978}%
  \BibitemOpen
  \bibfield  {author} {\bibinfo {author} {\bibfnamefont {E.~B.}\ \bibnamefont
  {Sonin}},\ }\href@noop {} {\bibfield  {journal} {\bibinfo  {journal} {Sov.
  Phys. JETP}\ }\textbf {\bibinfo {volume} {47}},\ \bibinfo {pages} {1091}
  (\bibinfo {year} {1978})}\BibitemShut {NoStop}%
\bibitem [{\citenamefont {Sonin}(2010)}]{SoninAP2010}%
  \BibitemOpen
  \bibfield  {author} {\bibinfo {author} {\bibfnamefont {E.~B.}\ \bibnamefont
  {Sonin}},\ }\href {\doibase 10.1080/00018731003739943} {\bibfield  {journal}
  {\bibinfo  {journal} {Adv. Phys.}\ }\textbf {\bibinfo {volume} {59}},\
  \bibinfo {pages} {181} (\bibinfo {year} {2010})}\BibitemShut {NoStop}%
\bibitem [{\citenamefont {Chen}\ and\ \citenamefont
  {Sigrist}(2014)}]{ChenPRB2014-2}%
  \BibitemOpen
  \bibfield  {author} {\bibinfo {author} {\bibfnamefont {W.}~\bibnamefont
  {Chen}}\ and\ \bibinfo {author} {\bibfnamefont {M.}~\bibnamefont {Sigrist}},\
  }\href {\doibase 10.1103/PhysRevB.89.024511} {\bibfield  {journal} {\bibinfo
  {journal} {Phys. Rev. B}\ }\textbf {\bibinfo {volume} {89}},\ \bibinfo
  {pages} {024511} (\bibinfo {year} {2014})}\BibitemShut {NoStop}%
\bibitem [{\citenamefont {Chen}\ \emph {et~al.}(2014)\citenamefont {Chen},
  \citenamefont {Kent}, \citenamefont {MacDonald},\ and\ \citenamefont
  {Sodemann}}]{ChenPRB2014}%
  \BibitemOpen
  \bibfield  {author} {\bibinfo {author} {\bibfnamefont {H.}~\bibnamefont
  {Chen}}, \bibinfo {author} {\bibfnamefont {A.~D.}\ \bibnamefont {Kent}},
  \bibinfo {author} {\bibfnamefont {A.~H.}\ \bibnamefont {MacDonald}}, \ and\
  \bibinfo {author} {\bibfnamefont {I.}~\bibnamefont {Sodemann}},\ }\href
  {\doibase 10.1103/PhysRevB.90.220401} {\bibfield  {journal} {\bibinfo
  {journal} {Phys. Rev. B}\ }\textbf {\bibinfo {volume} {90}},\ \bibinfo
  {pages} {220401(R)} (\bibinfo {year} {2014})}\BibitemShut {NoStop}%
\bibitem [{\citenamefont {Takei}\ and\ \citenamefont
  {Tserkovnyak}(2014)}]{TakeiPRL2014}%
  \BibitemOpen
  \bibfield  {author} {\bibinfo {author} {\bibfnamefont {S.}~\bibnamefont
  {Takei}}\ and\ \bibinfo {author} {\bibfnamefont {Y.}~\bibnamefont
  {Tserkovnyak}},\ }\href {\doibase 10.1103/PhysRevLett.112.227201} {\bibfield
  {journal} {\bibinfo  {journal} {Phys. Rev. Lett.}\ }\textbf {\bibinfo
  {volume} {112}},\ \bibinfo {pages} {227201} (\bibinfo {year}
  {2014})}\BibitemShut {NoStop}%
\bibitem [{\citenamefont {Takei}\ \emph {et~al.}(2014)\citenamefont {Takei},
  \citenamefont {Halperin}, \citenamefont {Yacoby},\ and\ \citenamefont
  {Tserkovnyak}}]{TakeiPRB2014}%
  \BibitemOpen
  \bibfield  {author} {\bibinfo {author} {\bibfnamefont {S.}~\bibnamefont
  {Takei}}, \bibinfo {author} {\bibfnamefont {B.~I.}\ \bibnamefont {Halperin}},
  \bibinfo {author} {\bibfnamefont {A.}~\bibnamefont {Yacoby}}, \ and\ \bibinfo
  {author} {\bibfnamefont {Y.}~\bibnamefont {Tserkovnyak}},\ }\href {\doibase
  10.1103/PhysRevB.90.094408} {\bibfield  {journal} {\bibinfo  {journal} {Phys.
  Rev. B}\ }\textbf {\bibinfo {volume} {90}},\ \bibinfo {pages} {094408}
  (\bibinfo {year} {2014})}\BibitemShut {NoStop}%
\bibitem [{\citenamefont {Chen}\ and\ \citenamefont
  {Sigrist}(2015)}]{ChenPRL2015}%
  \BibitemOpen
  \bibfield  {author} {\bibinfo {author} {\bibfnamefont {W.}~\bibnamefont
  {Chen}}\ and\ \bibinfo {author} {\bibfnamefont {M.}~\bibnamefont {Sigrist}},\
  }\href {\doibase 10.1103/PhysRevLett.114.157203} {\bibfield  {journal}
  {\bibinfo  {journal} {Phys. Rev. Lett.}\ }\textbf {\bibinfo {volume} {114}},\
  \bibinfo {pages} {157203} (\bibinfo {year} {2015})}\BibitemShut {NoStop}%
\bibitem [{\citenamefont {Kalmeyer}\ and\ \citenamefont
  {Laughlin}(1987)}]{KalmeyerPRL1987}%
  \BibitemOpen
  \bibfield  {author} {\bibinfo {author} {\bibfnamefont {V.}~\bibnamefont
  {Kalmeyer}}\ and\ \bibinfo {author} {\bibfnamefont {R.~B.}\ \bibnamefont
  {Laughlin}},\ }\href {\doibase 10.1103/PhysRevLett.59.2095} {\bibfield
  {journal} {\bibinfo  {journal} {Phys. Rev. Lett.}\ }\textbf {\bibinfo
  {volume} {59}},\ \bibinfo {pages} {2095} (\bibinfo {year}
  {1987})}\BibitemShut {NoStop}%
\bibitem [{\citenamefont {Kalmeyer}\ and\ \citenamefont
  {Laughlin}(1989)}]{KalmeyerPRB1989}%
  \BibitemOpen
  \bibfield  {author} {\bibinfo {author} {\bibfnamefont {V.}~\bibnamefont
  {Kalmeyer}}\ and\ \bibinfo {author} {\bibfnamefont {R.~B.}\ \bibnamefont
  {Laughlin}},\ }\href {\doibase 10.1103/PhysRevB.39.11879} {\bibfield
  {journal} {\bibinfo  {journal} {Phys. Rev. B}\ }\textbf {\bibinfo {volume}
  {39}},\ \bibinfo {pages} {11879} (\bibinfo {year} {1989})}\BibitemShut
  {NoStop}%
\bibitem [{\citenamefont {Wen}\ \emph {et~al.}(1989)\citenamefont {Wen},
  \citenamefont {Wilczek},\ and\ \citenamefont {Zee}}]{WenPRB1989}%
  \BibitemOpen
  \bibfield  {author} {\bibinfo {author} {\bibfnamefont {X.~G.}\ \bibnamefont
  {Wen}}, \bibinfo {author} {\bibfnamefont {F.}~\bibnamefont {Wilczek}}, \ and\
  \bibinfo {author} {\bibfnamefont {A.}~\bibnamefont {Zee}},\ }\href {\doibase
  10.1103/PhysRevB.39.11413} {\bibfield  {journal} {\bibinfo  {journal} {Phys.
  Rev. B}\ }\textbf {\bibinfo {volume} {39}},\ \bibinfo {pages} {11413}
  (\bibinfo {year} {1989})}\BibitemShut {NoStop}%
\bibitem [{\citenamefont {Meng}\ \emph {et~al.}(2015)\citenamefont {Meng},
  \citenamefont {Neupert}, \citenamefont {Greiter},\ and\ \citenamefont
  {Thomale}}]{MengPRB2015}%
  \BibitemOpen
  \bibfield  {author} {\bibinfo {author} {\bibfnamefont {T.}~\bibnamefont
  {Meng}}, \bibinfo {author} {\bibfnamefont {T.}~\bibnamefont {Neupert}},
  \bibinfo {author} {\bibfnamefont {M.}~\bibnamefont {Greiter}}, \ and\
  \bibinfo {author} {\bibfnamefont {R.}~\bibnamefont {Thomale}},\ }\href
  {\doibase 10.1103/PhysRevB.91.241106} {\bibfield  {journal} {\bibinfo
  {journal} {Phys. Rev. B}\ }\textbf {\bibinfo {volume} {91}},\ \bibinfo
  {pages} {241106(R)} (\bibinfo {year} {2015})}\BibitemShut {NoStop}%
\bibitem [{\citenamefont {Haldane}\ and\ \citenamefont
  {Arovas}(1995)}]{HaldanePRB1995}%
  \BibitemOpen
  \bibfield  {author} {\bibinfo {author} {\bibfnamefont {F.~D.~M.}\
  \bibnamefont {Haldane}}\ and\ \bibinfo {author} {\bibfnamefont {D.~P.}\
  \bibnamefont {Arovas}},\ }\href {\doibase 10.1103/PhysRevB.52.4223}
  {\bibfield  {journal} {\bibinfo  {journal} {Phys. Rev. B}\ }\textbf {\bibinfo
  {volume} {52}},\ \bibinfo {pages} {4223} (\bibinfo {year}
  {1995})}\BibitemShut {NoStop}%
\bibitem [{\citenamefont {Matsumoto}\ and\ \citenamefont
  {Murakami}(2011{\natexlab{a}})}]{MatsumotoPRL2011}%
  \BibitemOpen
  \bibfield  {author} {\bibinfo {author} {\bibfnamefont {R.}~\bibnamefont
  {Matsumoto}}\ and\ \bibinfo {author} {\bibfnamefont {S.}~\bibnamefont
  {Murakami}},\ }\href {\doibase 10.1103/PhysRevLett.106.197202} {\bibfield
  {journal} {\bibinfo  {journal} {Phys. Rev. Lett.}\ }\textbf {\bibinfo
  {volume} {106}},\ \bibinfo {pages} {197202} (\bibinfo {year}
  {2011}{\natexlab{a}})}\BibitemShut {NoStop}%
\bibitem [{\citenamefont {Matsumoto}\ and\ \citenamefont
  {Murakami}(2011{\natexlab{b}})}]{MatsumotoPRB2012}%
  \BibitemOpen
  \bibfield  {author} {\bibinfo {author} {\bibfnamefont {R.}~\bibnamefont
  {Matsumoto}}\ and\ \bibinfo {author} {\bibfnamefont {S.}~\bibnamefont
  {Murakami}},\ }\href {\doibase 10.1103/PhysRevB.84.184406} {\bibfield
  {journal} {\bibinfo  {journal} {Phys. Rev. B}\ }\textbf {\bibinfo {volume}
  {84}},\ \bibinfo {pages} {184406} (\bibinfo {year}
  {2011}{\natexlab{b}})}\BibitemShut {NoStop}%
\bibitem [{\citenamefont {Shindou}\ \emph
  {et~al.}(2013{\natexlab{a}})\citenamefont {Shindou}, \citenamefont {Ohe},
  \citenamefont {Matsumoto}, \citenamefont {Murakami},\ and\ \citenamefont
  {Saitoh}}]{ShindouPRB2013}%
  \BibitemOpen
  \bibfield  {author} {\bibinfo {author} {\bibfnamefont {R.}~\bibnamefont
  {Shindou}}, \bibinfo {author} {\bibfnamefont {J.-i.}\ \bibnamefont {Ohe}},
  \bibinfo {author} {\bibfnamefont {R.}~\bibnamefont {Matsumoto}}, \bibinfo
  {author} {\bibfnamefont {S.}~\bibnamefont {Murakami}}, \ and\ \bibinfo
  {author} {\bibfnamefont {E.}~\bibnamefont {Saitoh}},\ }\href {\doibase
  10.1103/PhysRevB.87.174402} {\bibfield  {journal} {\bibinfo  {journal} {Phys.
  Rev. B}\ }\textbf {\bibinfo {volume} {87}},\ \bibinfo {pages} {174402}
  (\bibinfo {year} {2013}{\natexlab{a}})}\BibitemShut {NoStop}%
\bibitem [{\citenamefont {Shindou}\ \emph
  {et~al.}(2013{\natexlab{b}})\citenamefont {Shindou}, \citenamefont
  {Matsumoto}, \citenamefont {Murakami},\ and\ \citenamefont
  {Ohe}}]{ShindouPRB2013-2}%
  \BibitemOpen
  \bibfield  {author} {\bibinfo {author} {\bibfnamefont {R.}~\bibnamefont
  {Shindou}}, \bibinfo {author} {\bibfnamefont {R.}~\bibnamefont {Matsumoto}},
  \bibinfo {author} {\bibfnamefont {S.}~\bibnamefont {Murakami}}, \ and\
  \bibinfo {author} {\bibfnamefont {J.-i.}\ \bibnamefont {Ohe}},\ }\href
  {\doibase 10.1103/PhysRevB.87.174427} {\bibfield  {journal} {\bibinfo
  {journal} {Phys. Rev. B}\ }\textbf {\bibinfo {volume} {87}},\ \bibinfo
  {pages} {174427} (\bibinfo {year} {2013}{\natexlab{b}})}\BibitemShut
  {NoStop}%
\bibitem [{\citenamefont {Zhang}\ \emph {et~al.}(2013)\citenamefont {Zhang},
  \citenamefont {Ren}, \citenamefont {Wang},\ and\ \citenamefont
  {Li}}]{ZhangPRB2013}%
  \BibitemOpen
  \bibfield  {author} {\bibinfo {author} {\bibfnamefont {L.}~\bibnamefont
  {Zhang}}, \bibinfo {author} {\bibfnamefont {J.}~\bibnamefont {Ren}}, \bibinfo
  {author} {\bibfnamefont {J.-S.}\ \bibnamefont {Wang}}, \ and\ \bibinfo
  {author} {\bibfnamefont {B.}~\bibnamefont {Li}},\ }\href {\doibase
  10.1103/PhysRevB.87.144101} {\bibfield  {journal} {\bibinfo  {journal} {Phys.
  Rev. B}\ }\textbf {\bibinfo {volume} {87}},\ \bibinfo {pages} {144101}
  (\bibinfo {year} {2013})}\BibitemShut {NoStop}%
\bibitem [{\citenamefont {Mook}\ \emph
  {et~al.}(2014{\natexlab{a}})\citenamefont {Mook}, \citenamefont {Henk},\ and\
  \citenamefont {Mertig}}]{MookPRB2014}%
  \BibitemOpen
  \bibfield  {author} {\bibinfo {author} {\bibfnamefont {A.}~\bibnamefont
  {Mook}}, \bibinfo {author} {\bibfnamefont {J.}~\bibnamefont {Henk}}, \ and\
  \bibinfo {author} {\bibfnamefont {I.}~\bibnamefont {Mertig}},\ }\href
  {\doibase 10.1103/PhysRevB.89.134409} {\bibfield  {journal} {\bibinfo
  {journal} {Phys. Rev. B}\ }\textbf {\bibinfo {volume} {89}},\ \bibinfo
  {pages} {134409} (\bibinfo {year} {2014}{\natexlab{a}})}\BibitemShut
  {NoStop}%
\bibitem [{\citenamefont {Mook}\ \emph
  {et~al.}(2014{\natexlab{b}})\citenamefont {Mook}, \citenamefont {Henk},\ and\
  \citenamefont {Mertig}}]{MookPRB2014-2}%
  \BibitemOpen
  \bibfield  {author} {\bibinfo {author} {\bibfnamefont {A.}~\bibnamefont
  {Mook}}, \bibinfo {author} {\bibfnamefont {J.}~\bibnamefont {Henk}}, \ and\
  \bibinfo {author} {\bibfnamefont {I.}~\bibnamefont {Mertig}},\ }\href
  {\doibase 10.1103/PhysRevB.90.024412} {\bibfield  {journal} {\bibinfo
  {journal} {Phys. Rev. B}\ }\textbf {\bibinfo {volume} {90}},\ \bibinfo
  {pages} {024412} (\bibinfo {year} {2014}{\natexlab{b}})}\BibitemShut
  {NoStop}%
\bibitem [{\citenamefont {Yakovenko}(1991)}]{YakovenkoPRB1991}%
  \BibitemOpen
  \bibfield  {author} {\bibinfo {author} {\bibfnamefont {V.~M.}\ \bibnamefont
  {Yakovenko}},\ }\href {\doibase 10.1103/PhysRevB.43.11353} {\bibfield
  {journal} {\bibinfo  {journal} {Phys. Rev. B}\ }\textbf {\bibinfo {volume}
  {43}},\ \bibinfo {pages} {11353} (\bibinfo {year} {1991})}\BibitemShut
  {NoStop}%
\bibitem [{\citenamefont {Klinovaja}\ and\ \citenamefont
  {Loss}(2013)}]{KlinovajaPRL2013}%
  \BibitemOpen
  \bibfield  {author} {\bibinfo {author} {\bibfnamefont {J.}~\bibnamefont
  {Klinovaja}}\ and\ \bibinfo {author} {\bibfnamefont {D.}~\bibnamefont
  {Loss}},\ }\href {\doibase 10.1103/PhysRevLett.111.196401} {\bibfield
  {journal} {\bibinfo  {journal} {Phys. Rev. Lett.}\ }\textbf {\bibinfo
  {volume} {111}},\ \bibinfo {pages} {196401} (\bibinfo {year}
  {2013})}\BibitemShut {NoStop}%
\bibitem [{\citenamefont {Kane}\ \emph {et~al.}(2002)\citenamefont {Kane},
  \citenamefont {Mukhopadhyay},\ and\ \citenamefont {Lubensky}}]{KanePRL2002}%
  \BibitemOpen
  \bibfield  {author} {\bibinfo {author} {\bibfnamefont {C.~L.}\ \bibnamefont
  {Kane}}, \bibinfo {author} {\bibfnamefont {R.}~\bibnamefont {Mukhopadhyay}},
  \ and\ \bibinfo {author} {\bibfnamefont {T.~C.}\ \bibnamefont {Lubensky}},\
  }\href {\doibase 10.1103/PhysRevLett.88.036401} {\bibfield  {journal}
  {\bibinfo  {journal} {Phys. Rev. Lett.}\ }\textbf {\bibinfo {volume} {88}},\
  \bibinfo {pages} {036401} (\bibinfo {year} {2002})}\BibitemShut {NoStop}%
\bibitem [{\citenamefont {Klinovaja}\ and\ \citenamefont
  {Tserkovnyak}(2014)}]{KlinovajaPRB2014}%
  \BibitemOpen
  \bibfield  {author} {\bibinfo {author} {\bibfnamefont {J.}~\bibnamefont
  {Klinovaja}}\ and\ \bibinfo {author} {\bibfnamefont {Y.}~\bibnamefont
  {Tserkovnyak}},\ }\href {\doibase 10.1103/PhysRevB.90.115426} {\bibfield
  {journal} {\bibinfo  {journal} {Phys. Rev. B}\ }\textbf {\bibinfo {volume}
  {90}},\ \bibinfo {pages} {115426} (\bibinfo {year} {2014})}\BibitemShut
  {NoStop}%
\bibitem [{\citenamefont {Jordan}\ and\ \citenamefont
  {Wigner}(1928)}]{JordanZP1928}%
  \BibitemOpen
  \bibfield  {author} {\bibinfo {author} {\bibfnamefont {P.}~\bibnamefont
  {Jordan}}\ and\ \bibinfo {author} {\bibfnamefont {E.}~\bibnamefont
  {Wigner}},\ }\href {\doibase 10.1007/BF01331938} {\bibfield  {journal}
  {\bibinfo  {journal} {Z. Phys.}\ }\textbf {\bibinfo {volume} {47}},\ \bibinfo
  {pages} {631} (\bibinfo {year} {1928})}\BibitemShut {NoStop}%
\bibitem [{\citenamefont {Galitski}(2010)}]{GalitskiPRB2010}%
  \BibitemOpen
  \bibfield  {author} {\bibinfo {author} {\bibfnamefont {V.}~\bibnamefont
  {Galitski}},\ }\href {\doibase 10.1103/PhysRevB.82.060411} {\bibfield
  {journal} {\bibinfo  {journal} {Phys. Rev. B}\ }\textbf {\bibinfo {volume}
  {82}},\ \bibinfo {pages} {060411} (\bibinfo {year} {2010})}\BibitemShut
  {NoStop}%
\bibitem [{\citenamefont {Micnas}\ \emph {et~al.}(1990)\citenamefont {Micnas},
  \citenamefont {Ranninger},\ and\ \citenamefont
  {Robaszkiewicz}}]{MicnasRMP1990}%
  \BibitemOpen
  \bibfield  {author} {\bibinfo {author} {\bibfnamefont {R.}~\bibnamefont
  {Micnas}}, \bibinfo {author} {\bibfnamefont {J.}~\bibnamefont {Ranninger}}, \
  and\ \bibinfo {author} {\bibfnamefont {S.}~\bibnamefont {Robaszkiewicz}},\
  }\href {\doibase 10.1103/RevModPhys.62.113} {\bibfield  {journal} {\bibinfo
  {journal} {Rev. Mod. Phys.}\ }\textbf {\bibinfo {volume} {62}},\ \bibinfo
  {pages} {113} (\bibinfo {year} {1990})}\BibitemShut {NoStop}%
\bibitem [{Note1()}]{Note1}%
  \BibitemOpen
  \bibinfo {note} {We do not lose any generality by choosing the sign of the
  exchange interaction, $J > 0$ here, for the sign can be flipped by rotating
  every other spin by $\pi $ around the $z$ axis.}\BibitemShut {Stop}%
\bibitem [{Note2()}]{Note2}%
  \BibitemOpen
  \bibinfo {note} {The long-range interchain interactions may be treated within
  the framework of the coupled Luttinger liquids~\cite {*[][{, and references
  therein.}] GiamarchiCR2004}, but it is beyond the scope of our
  work.}\BibitemShut {Stop}%
\bibitem [{Note3()}]{Note3}%
  \BibitemOpen
  \bibinfo {note} {The comparison of two tuples, $(k, \beta )$ and $(m, \alpha
  )$, representing two spin chains is determined by their orders in a
  serialization of spin chains. One can take an alternative JW transformation
  by restricting $\beta $ to $\alpha $ in the product and introducing a new
  additional Pauli-matrix vector for maintaining anticommutation between the
  two layers.}\BibitemShut {Stop}%
\bibitem [{Note4()}]{Note4}%
  \BibitemOpen
  \bibinfo {note} {The Hartree-Fock decoupling is taken only for the interchain
  interaction, not for the Ising interlayer interaction.}\BibitemShut {Stop}%
\bibitem [{Note5()}]{Note5}%
  \BibitemOpen
  \bibinfo {note} {In the weak-attraction limit, $u \ll t_x \protect \tmspace
  +\thinmuskip {.1667em} , t_y$, the superconducting gap is given by $\Delta
  \sim t \protect \sqrt {\rho (2 - \rho )} / \protect \qopname \relax o{sinh}(t
  / u)$, where $t \equiv t_x + t_y$ is the bandwidth and $\rho $ is the number
  of electrons per site~\cite {MicnasRMP1990}.}\BibitemShut {Stop}%
\bibitem [{Note6()}]{Note6}%
  \BibitemOpen
  \bibinfo {note} {For the strong DM interactions comparing to the interchain
  coupling, $0 < K \ll \phi J$, the mean field for the interchain tunneling is
  exponentially small, $\chi \sim \protect \qopname \relax o{exp}(- 4 \pi \phi
  J / K)$.}\BibitemShut {Stop}%
\bibitem [{\citenamefont {Qi}\ and\ \citenamefont {Zhang}(2011)}]{QiRMP2011}%
  \BibitemOpen
  \bibfield  {author} {\bibinfo {author} {\bibfnamefont {X.-L.}\ \bibnamefont
  {Qi}}\ and\ \bibinfo {author} {\bibfnamefont {S.-C.}\ \bibnamefont {Zhang}},\
  }\href {\doibase 10.1103/RevModPhys.83.1057} {\bibfield  {journal} {\bibinfo
  {journal} {Rev. Mod. Phys.}\ }\textbf {\bibinfo {volume} {83}},\ \bibinfo
  {pages} {1057} (\bibinfo {year} {2011})}\BibitemShut {NoStop}%
\bibitem [{\citenamefont {Breunig}\ \emph {et~al.}(2013)\citenamefont
  {Breunig}, \citenamefont {Garst}, \citenamefont {Sela}, \citenamefont
  {Buldmann}, \citenamefont {Becker}, \citenamefont {Bohat\'y}, \citenamefont
  {M\"uller},\ and\ \citenamefont {Lorenz}}]{BreunigPRL2013}%
  \BibitemOpen
  \bibfield  {author} {\bibinfo {author} {\bibfnamefont {O.}~\bibnamefont
  {Breunig}}, \bibinfo {author} {\bibfnamefont {M.}~\bibnamefont {Garst}},
  \bibinfo {author} {\bibfnamefont {E.}~\bibnamefont {Sela}}, \bibinfo {author}
  {\bibfnamefont {B.}~\bibnamefont {Buldmann}}, \bibinfo {author}
  {\bibfnamefont {P.}~\bibnamefont {Becker}}, \bibinfo {author} {\bibfnamefont
  {L.}~\bibnamefont {Bohat\'y}}, \bibinfo {author} {\bibfnamefont
  {R.}~\bibnamefont {M\"uller}}, \ and\ \bibinfo {author} {\bibfnamefont
  {T.}~\bibnamefont {Lorenz}},\ }\href {\doibase
  10.1103/PhysRevLett.111.187202} {\bibfield  {journal} {\bibinfo  {journal}
  {Phys. Rev. Lett.}\ }\textbf {\bibinfo {volume} {111}},\ \bibinfo {pages}
  {187202} (\bibinfo {year} {2013})}\BibitemShut {NoStop}%
\bibitem [{\citenamefont {Goovaerts}\ \emph {et~al.}(1984)\citenamefont
  {Goovaerts}, \citenamefont {De~Raedt},\ and\ \citenamefont
  {Schoemaker}}]{GoovaertsPRL1984}%
  \BibitemOpen
  \bibfield  {author} {\bibinfo {author} {\bibfnamefont {E.}~\bibnamefont
  {Goovaerts}}, \bibinfo {author} {\bibfnamefont {H.}~\bibnamefont {De~Raedt}},
  \ and\ \bibinfo {author} {\bibfnamefont {D.}~\bibnamefont {Schoemaker}},\
  }\href {\doibase 10.1103/PhysRevLett.52.1649} {\bibfield  {journal} {\bibinfo
   {journal} {Phys. Rev. Lett.}\ }\textbf {\bibinfo {volume} {52}},\ \bibinfo
  {pages} {1649} (\bibinfo {year} {1984})}\BibitemShut {NoStop}%
\bibitem [{\citenamefont {Mikeska}\ and\ \citenamefont
  {Kolezhuk}(2004)}]{Mikeska2004}%
  \BibitemOpen
  \bibfield  {author} {\bibinfo {author} {\bibfnamefont {H.-J.}\ \bibnamefont
  {Mikeska}}\ and\ \bibinfo {author} {\bibfnamefont {A.}~\bibnamefont
  {Kolezhuk}},\ }\enquote {\bibinfo {title} {One-dimensional magnetism},}\ in\
  \href {http://dx.doi.org/10.1007/BFb0119591} {\emph {\bibinfo {booktitle}
  {Quantum Magnetism}}},\ \bibinfo {editor} {edited by\ \bibinfo {editor}
  {\bibfnamefont {U.}~\bibnamefont {Schollw{\"o}ck}}, \bibinfo {editor}
  {\bibfnamefont {J.}~\bibnamefont {Richter}}, \bibinfo {editor} {\bibfnamefont
  {D.}~\bibnamefont {Farnell}}, \ and\ \bibinfo {editor} {\bibfnamefont
  {R.}~\bibnamefont {Bishop}}}\ (\bibinfo  {publisher} {Springer Berlin
  Heidelberg},\ \bibinfo {year} {2004})\ \bibinfo {type}
  {10.1007/bfb0119591}~\bibinfo {chapter} {1}, pp.\ \bibinfo {pages}
  {1--83}\BibitemShut {NoStop}%
\bibitem [{\citenamefont {Gritsev}\ \emph {et~al.}(2004)\citenamefont
  {Gritsev}, \citenamefont {Normand},\ and\ \citenamefont
  {Baeriswyl}}]{GritsevPRB2004}%
  \BibitemOpen
  \bibfield  {author} {\bibinfo {author} {\bibfnamefont {V.}~\bibnamefont
  {Gritsev}}, \bibinfo {author} {\bibfnamefont {B.}~\bibnamefont {Normand}}, \
  and\ \bibinfo {author} {\bibfnamefont {D.}~\bibnamefont {Baeriswyl}},\ }\href
  {\doibase 10.1103/PhysRevB.69.094431} {\bibfield  {journal} {\bibinfo
  {journal} {Phys. Rev. B}\ }\textbf {\bibinfo {volume} {69}},\ \bibinfo
  {pages} {094431} (\bibinfo {year} {2004})}\BibitemShut {NoStop}%
\bibitem [{\citenamefont {Capponi}\ \emph {et~al.}(2013)\citenamefont
  {Capponi}, \citenamefont {Lecheminant},\ and\ \citenamefont
  {Moliner}}]{CapponiPRB2013}%
  \BibitemOpen
  \bibfield  {author} {\bibinfo {author} {\bibfnamefont {S.}~\bibnamefont
  {Capponi}}, \bibinfo {author} {\bibfnamefont {P.}~\bibnamefont
  {Lecheminant}}, \ and\ \bibinfo {author} {\bibfnamefont {M.}~\bibnamefont
  {Moliner}},\ }\href {\doibase 10.1103/PhysRevB.88.075132} {\bibfield
  {journal} {\bibinfo  {journal} {Phys. Rev. B}\ }\textbf {\bibinfo {volume}
  {88}},\ \bibinfo {pages} {075132} (\bibinfo {year} {2013})}\BibitemShut
  {NoStop}%
\bibitem [{\citenamefont {Bruus}\ and\ \citenamefont {Flesberg}(2004)}]{Bruus}%
  \BibitemOpen
  \bibfield  {author} {\bibinfo {author} {\bibfnamefont {H.}~\bibnamefont
  {Bruus}}\ and\ \bibinfo {author} {\bibfnamefont {K.}~\bibnamefont
  {Flesberg}},\ }\href@noop {} {\emph {\bibinfo {title} {Many-body Quantum
  Theory in Condensed Matter Physics}}}\ (\bibinfo  {publisher} {Oxford
  University Press, Oxford},\ \bibinfo {year} {2004})\BibitemShut {NoStop}%
\bibitem [{Note7()}]{Note7}%
  \BibitemOpen
  \bibinfo {note} {There is the third mean field, $\DOTSB \sum@ \slimits@ _{n,
  m, \alpha } \protect \langle f^\dagger _{n, m, \alpha } f_{n+1, m+1, \alpha }
  \protect \rangle $, but it is omitted from the analysis because its effects
  are similar to those of $\chi $ and thus do not change the qualitative
  property of the system.}\BibitemShut {Stop}%
\bibitem [{\citenamefont {Sinova}\ \emph {et~al.}(2015)\citenamefont {Sinova},
  \citenamefont {Valenzuela}, \citenamefont {Wunderlich}, \citenamefont
  {Back},\ and\ \citenamefont {Jungwirth}}]{SinovaRMP2015}%
  \BibitemOpen
  \bibfield  {author} {\bibinfo {author} {\bibfnamefont {J.}~\bibnamefont
  {Sinova}}, \bibinfo {author} {\bibfnamefont {S.~O.}\ \bibnamefont
  {Valenzuela}}, \bibinfo {author} {\bibfnamefont {J.}~\bibnamefont
  {Wunderlich}}, \bibinfo {author} {\bibfnamefont {C.~H.}\ \bibnamefont
  {Back}}, \ and\ \bibinfo {author} {\bibfnamefont {T.}~\bibnamefont
  {Jungwirth}},\ }\href {\doibase 10.1103/RevModPhys.87.1213} {\bibfield
  {journal} {\bibinfo  {journal} {Rev. Mod. Phys.}\ }\textbf {\bibinfo {volume}
  {87}},\ \bibinfo {pages} {1213} (\bibinfo {year} {2015})}\BibitemShut
  {NoStop}%
\bibitem [{\citenamefont {Giamarchi}(2004)}]{GiamarchiCR2004}%
  \BibitemOpen
  \bibfield  {author} {\bibinfo {author} {\bibfnamefont {T.}~\bibnamefont
  {Giamarchi}},\ }\href {\doibase 10.1021/cr030647c} {\bibfield  {journal}
  {\bibinfo  {journal} {Chem. Rev.}\ }\textbf {\bibinfo {volume} {104}},\
  \bibinfo {pages} {5037} (\bibinfo {year} {2004})}\BibitemShut {NoStop}%
\end{thebibliography}%

\end{document}